\documentclass[superscriptaddress,twocolumn,amsmath,amssymb,aps,pre,english]{revtex4-2}

\usepackage{dcolumn}
\usepackage{bm}
\usepackage{natbib}
\usepackage{ulem} 
\usepackage{changebar}
\usepackage[T1]{fontenc}
\usepackage{inputenc}
\usepackage{upgreek,babel,calc,mathrsfs}
\usepackage{amsmath,amssymb,graphicx,xcolor}
\usepackage[export]{adjustbox}
\usepackage[unicode=true]{hyperref}
\newcommand{\mathd}{\mathrm{d}}
\newcommand{\tmop}[1]{\mathrm{#1}}
\newcommand{\acrit}{\mathcal{A}_\mathrm{c}}

\begin{document}

\title{Phase separation and critical size in molecular sorting}

\affiliation{Institute of Condensed Matter Physics and Complex Systems,
Department of Applied Science and Technology, Politecnico di Torino,
Corso Duca degli Abruzzi 24, 10129 Torino, Italy}
\author{Elisa Floris}
\thanks{These authors contributed equally.}
\affiliation{Institute of Condensed Matter Physics and Complex Systems,
Department of Applied Science and Technology, Politecnico di Torino,
Corso Duca degli Abruzzi 24, 10129 Torino, Italy}
\author{Andrea Piras}
\thanks{These authors contributed equally.}
\affiliation{Italian Institute for Genomic Medicine (IIGM) and Candiolo Cancer Institute IRCCS, str.~prov.~142, km 3.95, Candiolo (TO) 10060, Italy}
\author{Francesco Saverio Pezzicoli}
\thanks{These authors contributed equally.}
\affiliation{Laboratoire Interdisciplinaire des Sciences du Numérique (LISN), Université Paris-Saclay, Gif-sur-Yvette, Île-de-France, France}
\author{Marco Zamparo}
\affiliation{Dipartimento di Fisica, Università degli Studi di Bari, via Amendola 173,
70126 Bari, Italy}
\affiliation{Istituto Nazionale di Fisica Nucleare (INFN), Italy}
\author{Luca Dall'Asta}
\email{luca.dallasta@polito.it}
\affiliation{Institute of Condensed Matter Physics and Complex Systems,
Department of Applied Science and Technology,
Politecnico di Torino, Corso Duca degli Abruzzi 24, 10129 Torino, Italy}
\affiliation{Collegio Carlo Alberto, Piazza Arbarello 8, 10122, Torino, Italy}
\affiliation{Istituto Nazionale di Fisica Nucleare (INFN), Italy}
\affiliation{Italian Institute for Genomic Medicine (IIGM) and Candiolo Cancer Institute IRCCS, str.~prov.~142, km 3.95, Candiolo (TO) 10060, Italy}
\author{Andrea Gamba}
\email{andrea.gamba@polito.it}
\affiliation{Institute of Condensed Matter Physics and Complex Systems,
Department of Applied Science and Technology, Politecnico di Torino,
Corso Duca degli Abruzzi 24, 10129 Torino, Italy}
\affiliation{Istituto Nazionale di Fisica Nucleare (INFN), Italy}
\affiliation{Italian Institute for Genomic Medicine (IIGM) and Candiolo Cancer Institute IRCCS, str.~prov.~142, km 3.95, Candiolo (TO) 10060, Italy}

\begin{abstract}
Molecular sorting is a fundamental process 
that allows eukaryotic
cells to distill and concentrate specific chemical factors in appropriate cell membrane subregions, thus endowing them with different chemical identities and functional properties. A 
phenomenological theory of this molecular distillation process 
has recently been proposed~\cite{ZVS+21},
based on the idea
that molecular sorting emerges from the combination of:
a)~phase-separation-driven
formation of sorting domains,  
and 
b)~domain-induced membrane bending, leading to the 
production
of submicrometric lipid vesicles enriched in the sorted molecules. In this framework, a natural parameter controlling the efficiency of molecular distillation is
the 
critical size of phase-separated domains.
In the experiments, sorting domains 
appear
to fall into two classes: unproductive domains, characterized by short lifetimes and low probability of 
extraction,
and productive domains, that evolve into vesicles 
that ultimately 
detach
from the membrane system. It 
is tempting 
to link these two classes to the different fates predicted by classical 
phase separation
theory for subcritical and supercritical phase-separated domains. Here, we discuss the implication of this picture in the framework of the
previously introduced phenomenological theory of molecular sorting. 
Several predictions of the  
theory are verified by  numerical simulations of a lattice-gas model. 
Sorting is observed to
be most efficient when the number of sorting domains is close to a minimum. 
To help in the
analysis of experimental data, 
an operational definition of 
the 
critical size
of sorting domains
is 
proposed.
Comparison with experimental
results shows that the statistical properties of productive/unproductive domains inferred from experimental
data are in agreement with those predicted from numerical simulations of the model, compatibly with the hypothesis
that molecular sorting 
is driven by a phase separation
process.
\end{abstract}

\keywords{Phase separation, protein sorting, nucleation theory}

\maketitle

\section{Introduction}

Molecular sorting 
is a major process responsible for the organization of cellular matter in eukaryotic cells~\cite{MN08}. This highly complex task is accomplished by selectively concentrating and distilling specific proteins and lipids that dwell on the plasma membrane and on the membranes of inner cellular bodies into submicrometric lipid vesicles. Once formed, these vesicles detach from the membrane 
and are
subsequently delivered to their appropriate destinations. 
It has recently been proposed that molecular sorting may emerge from the combination of two fundamental physical processes~\cite{ZVS+21}:
a) phase separation of specific molecules into localized sorting domains, and b) 
domain-induced membrane bending,
leading to the formation of 
vesicles
constitutively enriched in the biochemical factors of the engulfed domains,
thus resulting in a natural distillation process.
In 
the proposed abstract
model of 
the process, molecules arriving on a membrane region 
can
laterally diffuse and aggregate into localized domains, whose formation and growth occurs through the typical 
stages of 
phase 
separation:
after the initial nucleation stage, in the case of low supersaturation, the growth of domains is mainly governed by the absorption of freely diffusing molecules. 
One of the main predictions of the classical theory of phase separation is that
a {\it critical size}~$A_{\rm c}$ has to be reached in order for domains to survive and continue to grow irreversibly to larger and larger scales~\cite{LP81,Sle09}. 
In the present theory of molecular distillation such
domains are extracted once they reach a characteristic size~$A_E\gg A_\mathrm{c}$, 
determined by the physical and biomolecular  
processes that induce 
membrane bending and 
vesicle formation.
In the presence of a constant
flux of incoming molecules,
the membrane system selforganizes in
a driven non-equilibrium stationary state, which 
can
be 
seen
as a realization in Nature of the classical Szilard's model of 
droplet
formation~\cite{Far27,SR97,Sle09}. 

Phase separation phenomena are emerging as  central drivers of  the selforganization 
of cell structures~\cite{BBH18,HWJ14,LPR21,GCT+05,FPA+21},
and the idea that 
phase separation 
is
an essential step for molecular sorting 
is increasingly finding support in recent studies~\cite{BKH+21,KK22,DKW+21,ZZ20,LSD22}. 
As advances in live-cell imaging have 
enabled more accurate observations in real time, a striking heterogeneity in 
domain growth
kinetics 
has
emerged, and several approaches to unambiguously classify different dynamic populations 
have been
proposed~\cite{LMY+09,AAM+13,KSL+17,GCH+14,HSU+20,WCM+20}. In~the experiments, a crucial parameter used to 
describe 
the 
sorting 
process 
is the 
lifetime of a 
sorting domain.
It has been recently shown that the lifetime of a 
sorting domain
is 
related to 
the domain stability, which in its turn depends on the number of molecules  
contained in the 
domain, and thus 
on the domain size~\cite{LLN+19}. 
It is therefore tempting to relate the 
existence, in the context of phase separation,
of a critical size for domain growth, to the
observation that
sorting domains on 
cell membranes
can undergo qualitatively different final fates.
As a matter of fact, sorting domains are commonly classified 
in 
two groups:
{\it productive} domains, if their growth eventually terminates in the nucleation of a 
vesicle which 
is ultimately detached
from the membrane, and {\it unproductive} (or abortive) domains which, instead, progressively dismantle and are
ultimately dissolved~\cite{EBV+04,AAM+13,WCM+20}.
It seems natural to interpret this distinction in the context of classical nucleation theory, where the fate of a domain results from the balance between bulk stabilization and the propensity to dismantle along the domain boundary, which in its turn is controlled by the value of a characteristic boundary tension~\cite{LP81,GKL+09,FPA+21}. As a result, circular domains (that minimize the boundary perimeter) are favored, subcritical domains (having size $A<A_\mathrm{c}$) have short lifetimes and a low probability of reaching the extraction size $A_E$, while supercritical domains have a high probability of being ultimately extracted.
Here we discuss the implications of this picture in the framework of the phenomenological theory of molecular 
sorting
introduced in Ref.~\citenum{ZVS+21}.
Several predictions of the phenomenological theory are verified by extensive numerical simulations of a 
lattice-gas model.
To help in the analysis of experimental data, we introduce an operational definition of critical size, and discuss its relation to recently introduced methods for the classification of domain formation events into productive and unproductive classes~\cite{WCM+20}. The~operational definition is used here to compare the predictions of our phenomenological theory of molecular sorting to experiments on 
the formation of productive and unproductive
clathrin-coated pits at the plasma membrane. However, the proposed framework is more general, and we expect that it can turn useful in the interpretation of experiments on molecular sorting at different membrane regions, such as sorting endosomes, or the Golgi complex. A~direct comparison with experimental results shows that
the statistical properties of productive/unproductive domains inferred from experimental data are in good qualitative agreement with those emerging from simulations performed in
some specific parameter regions. These results hint at a central role of phase separation, and of the related notions of boundary tension and critical size, in the processes of molecular sorting 
that control 
the establishment and maintenance of distinct chemical identities 
on
cell membranes.

\section{Phenomenological theory}\label{sec:pheno}

We briefly summarize here the phenomenological theory of phase-separation-driven molecular sorting introduced in Ref.~\cite{ZVS+21}, and set up a convenient notation in view of the present discussion. The theory is based on the following non-equilibrium steady-state picture: a 
constant flux $\phi$ of ``sortable'' cargo
molecules is deposited on the lipid membrane; each molecule occupies a
characteristic area~$A_0$ on the membrane, diffuses laterally, and can
aggregate into sorting domains with the help of a pool of specialized 
auxiliary molecules, which 
sustain ``active'' domain formation by triggering
localized positive feedback loops~\cite{ZCT+15,FPA+21,GCT+05}, and/or ``passive'' aggregation, driven by 
weak attractive intermolecular interactions~\cite{LPR20,BBH18}. Since domain formation is characterized by competing effects, according to classical nucleation theory, a critical 
size~$A_{\rm c}$ is required for a domain to continue to grow irreversibly and avoid decay~\cite{BD35,Zel43,LP81}. 
Once formed, sorting domains coarsen due to the incoming flux of laterally diffusing
molecules, and are eventually extracted from the membrane in the form of lipid vesicles
of characteristic area $A_E = mA_0$.
It~follows that the growing domains coexist with a continuously repleted two-dimensional ``gas'' of laterally
diffusing molecules in a statistically stationary state. 

If we consider a 
region of linear size
$L$ 
of the order of the average interdomain half distance, centered around a 
growing supercritical domain of approximately circular shape and radius~$R$, 
the quasi-static profile $n_R(r)$ of the density of the gas of freely diffusing molecules in the proximity of  the domain
can be
approximately
obtained by solving 
a Laplace equation with 
Dirichlet boundary conditions 
$n_R(R)= n_0$ and $n_R(L)= \bar{n}$, obtaining
\begin{equation}
n_R(r) = n_0+ \frac{\log(r/R)}{\log(L/R)} \Delta n, \label{eq:densprofile}
\end{equation}
where $r\geq R$ 
denotes  the distance from the domain center, and $\Delta n=\bar{n} -n_0$. 
Domain growth is 
induced by
the flux $\Phi_A$ of molecules from the gas to the domain, which can be calculated by integrating the flux density $-D\nabla n_R(r)$ across the boundary of the domain of size $A=\pi R^2$, obtaining
\begin{equation}
 \Phi_A  =\frac{4 \pi D \Delta n}{\log(A_L/A)}
\end{equation}
where $D$ is the lateral  diffusivity of the molecules.
This formula implies that the domain will grow according to the dynamic equation
\begin{equation}
\dot{A}=\frac{4 \pi A_0 D \Delta n}{\log(A_L/A)}.\label{eq:dAdt}
\end{equation}
In a membrane system where sorting domains may be assumed to be approximately evenly distributed, the statistics of supercritical domains can be conveniently described in terms of the number density $N(t,A)\,\mathd A$, giving 
the average number per unit membrane area of supercritical domains with size comprised between $A$ and $A+\mathd A$.
Since the effects of random fluctuations can be approximately neglected in the case of supercritical domains, $N(A,t)$ satisfies the continuity equation
\begin{equation}
    \frac{\partial N}{\partial t} + \frac{\partial}{\partial A} (\dot A N)+\gamma (A)N=0,
    \label{eq:smo}
\end{equation}
where 
the rate of removal of 
domains of 
size $A$ 
from the system is 
$\gamma (A)=0$ for $A < A_{E}$, and $\gamma (A)=\gamma _0>0$ for $A > A_{E}$.
The stationary solution of Eq.~\eqref{eq:smo}, 
\begin{equation}\label{solutionSmolu}
    N_{\rm st}(A) = \frac{J \log{\left(A_L/A\right)}}{4 \pi D \Delta n} \exp \left [ -\int_{A_{\rm c}}^{A}{\frac{\gamma (a) \log{\left(A_L/a\right)}}{4 \pi A_0 D \Delta n} \mathd a} \right ]
\end{equation}
has
a universal logarithmic behavior for  $A<A_E$.
The normalization constant $J$ can be determined from the steady state condition
\begin{equation}
\phi  = \int_{A_\mathrm{c}}^\infty \Phi_A N_{\rm st}(A)\, \mathd A 
\simeq JA_E
\end{equation}
for large $\gamma_0$ and $A_E \gg A_{\rm c}$.
Assuming that the incoming flux $\phi$ of molecules is evenly distributed 
in average
among all available supercritical sorting domains, and neglecting logarithmic corrections, the 
average
number of supercritical domains per unit area is given by
\begin{equation}
 \bar{N}_d  \sim  \frac{\phi}{\Phi_A} \sim
  \frac{\phi}{D \Delta n}.  \label{eq:phioverDetc}
\end{equation}
Numerical observations suggest that faster responses of the membrane system to changing environmental conditions are related to shorter residence times of the sorted molecules on the membrane in the steady-state~\cite{ZVS+21}. It~is therefore interesting to investigate under which parametric conditions this residence time can be minimized.
From the moment of insertion to the moment of extraction, molecules spend an average time $\bar{T}_f$ diffusing 
freely 
and an average time $\bar{T}_d$ attached to 
supercritical
sorting domains. 
In principle, for 
the
molecules 
that aggregate
in the initial stage of 
the
domain formation process, when the 
domain is still subcritical,
one should also consider the time spent in the subcritical stage, but this is  generally negligible if the critical size is small.

In the following, repeated use 
will be made of a general steady-state relation, which applies to 
open systems in a driven non-equilibrium stationary state~\cite{ZAG19}: the average density of molecules in the system is given by the product of the average density flux of molecules (entering or leaving the system) and the average time that a molecule spends in the system. According to this general relation, the steady-state average density 
of molecules that are freely diffusing as a two-dimensional gas on the membrane is
\begin{eqnarray}
  \bar{n} & = & \phi \,\bar{T}_f.  \label{eq:dens}
\end{eqnarray}
The same steady-state relation can be applied to
the 
average density $\bar{N}_d$ of
supercritical
domains that are generated and ultimately extracted from the membrane, giving
\begin{eqnarray}
  \bar{N}_d & = & 
  \frac{{\rm d} \bar{N}_d}{{\rm d} t}\, \bar{T}_d \;=\;
  \frac{\phi }{m}\,\bar{T}_d. \label{eq:nd}
\end{eqnarray}
On the other hand,  since each new domain starts its aggregation process from
the encounter of two freely diffusing molecules, one can write (see also 
App.~\ref{app:islandmodel}) 
\begin{equation}
  \frac{\mathd \bar{N}_d}{\mathd t}  =  CD\,\bar{n}^2,  \label{eq:cdn2}
\end{equation}
where $C$ is a 
dimensionless
proportionality constant 
 measuring
 the strength of the effective interaction that keeps 
 molecules
together in a sorting domain. 
Combining (\ref{eq:dens}), (\ref{eq:nd}) and (\ref{eq:cdn2}), the following steady-state relations are obtained:
\begin{align}
\bar{n} & \sim \left(\frac{\phi}{m\, C D} \right)^{1/2},\\
\bar{T}_f & = \frac{\bar{n}}{\phi} \sim \left( m\, C\, D \phi \right)^{-1/2}.
\end{align}
For approximately absorbing domains, $n_0 \ll \Delta n$ and  $\Delta n \sim \bar n$, therefore (\ref{eq:phioverDetc}), (\ref{eq:nd}) and (\ref{eq:cdn2}) give:
\begin{align}
\bar{N}_d & \sim \left( \frac{m\, C \phi }{D}\right)^{1/2} \sim\; m\, C\,\bar{n}\,,
\label{eq:enned}
\\
\bar{T}_d & \sim \frac{C\, m^2\,\bar{n}}{\phi}  \sim \left( \frac{C\, m^3}{D\, \phi}\right)^{1/2}.
\end{align}
The average time spent by molecules in the system is approximately 
$\bar{T}=\bar{T}_d + \bar{T}_f$, which is minimum for
\begin{equation}
\bar{T}_f + \bar{T}_d  = \bar{T}_{\rm opt}  \sim
 \left( \frac{m}{D\, \phi} \right)^{1 / 2}. 
 \end{equation}
The optimal value $\bar{T}_{\rm opt}$  is obtained for 
\begin{equation}
C\;=\;C_{\rm opt} \sim \frac{1}{m^{2}}\,.
\label{eq:coptimal}
\end{equation} 
For this value, the average number densities of gas molecules and of supercritical domains are:
\begin{equation}
\bar{n}_{\rm opt} \sim \left(\frac{\phi A_E}{D A_0}\right)^{1/2}, \quad \bar{N}_{d, {\rm opt}}  \sim
 \left( \frac{\phi}{m\,D} \right)^{1 / 2}.
\end{equation}

\section{Numerical validation }\label{sec:validation}
\begin{figure*}[htb]
\begin{center}
\includegraphics[width=1\textwidth]{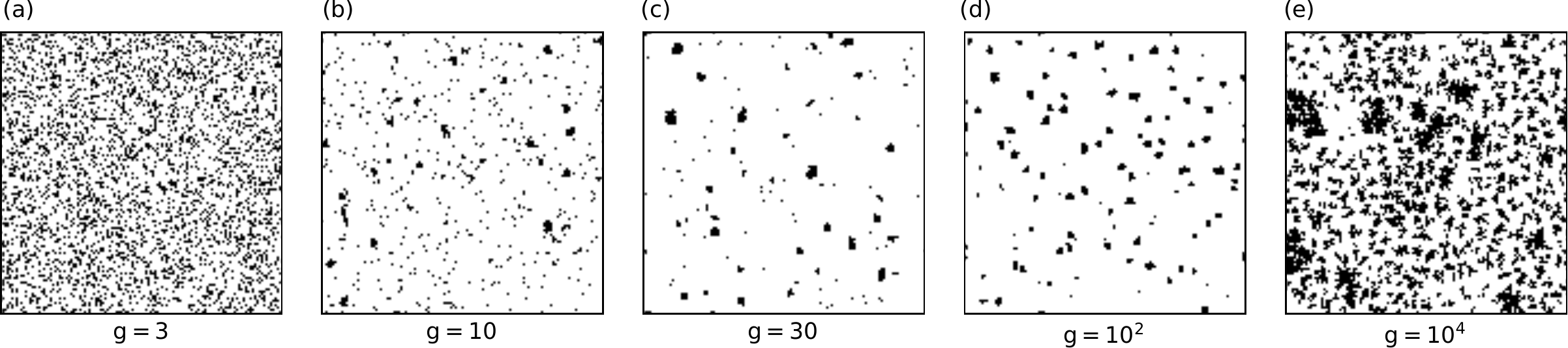}
 \end{center}
 \vspace{-.3cm}
 \caption{Snapshots of configurations 
 of the lattice-gas model 
 of molecular sorting
 for a system
 of $400^2$ sites
 in the steady state,
 with
 incoming flux
 $\phi/k_D = 10^{-6}$ 
 and increasing values of the 
 interaction
 strength~$g$ (from left to right). 
 In the central panel the interaction strength is close to the optimal value $g_\mathrm{opt}=31$.}
 \label{fig:snapshots}
\end{figure*}
In a minimal lattice-gas model of the distillation process, the
lipid membrane is modelled as a two-dimensional square lattice with periodic boundary conditions, where each site can be occupied by a single molecule at most~\cite{ZVS+21}. The system evolves according to a Markov process consisting of the following three elementary events: 
1) {\it insertion}: molecules from an infinite reservoir arrive and are inserted on empty sites with rate $k_I$; 
2) {\it diffusion and aggregation}: molecules can perform diffusive jumps to an empty neighboring site with rate $k_D/g^{N_\mathrm{nn}}$, where $g > 1$ is a dimensionless parameter 
representing
the interaction strength, and $N_\mathrm{nn}$ is the number of neighboring molecules of the hopping molecule before the jump occurs; 
3) {\it extraction}: molecules are extracted from the system 
by simultaneously removing all connected clusters of molecules that contain a completely filled square of 
size $m$.
In what follows, $A_0=1$, i.e.
areas 
are
measured 
as numbers of lattice sites, and 
$m=10^2$.
In 
every simulation, 
the system 
is allowed to relax to the steady state before starting the collection of 
relevant statistical data.

One of the main observations of Ref.~\citenum{ZVS+21} is that both the average permanence time $\bar{T}$ of sorted molecules on the membrane system and the average molecule density~$\rho$ in the steady state are minimal in an intermediate, optimal range of values of the interaction strength~$g$, where the molecular distillation process is most efficient.
Snapshots of the simulations taken in the steady state show the typical behavior of the system 
both inside and outside of this optimal range~(Fig.~\ref{fig:snapshots}).
For low interaction 
strength, molecular crowding accompanied by
a hectic formation of 
small 
short-lived domains
is observed
(Fig.~~\ref{fig:snapshots}(a)). As
the interaction strength increases,
the density of freely diffusing molecules decreases 
(Fig.~\ref{fig:snapshots}(b-d)). 
Consistently with the predictions of the phenomenological theory, the molecular density $\rho$ and residence time~$\bar{T}$ are lower in this intermediate range, and reach a minimum in correspondence with the optimal value of the interaction strength~$g$ (Ref.~\citenum{ZVS+21} and Fig.~\ref{fig:snapshots}(c)).
When the interaction strength becomes much larger than its optimal value, the gas of free molecules is strongly depleted, and the system enters into a regime of domain crowding (Fig.~\ref{fig:snapshots}(d)). Here, a large number of sorting domains shares the incoming molecular flux, the growth of each sorting domain is slowed down, and the efficiency of the distillation process is impaired, as both the molecular density and molecular residence time are much larger than in the optimal region.
For very high values of the microscopic interaction strength~$g$, the formation of highly irregular domains of the type predicted by 
the theory of diffusion-limited aggregation~\cite{BS95} is observed (Fig.~\ref{fig:snapshots}(e)). This latter regime is unlikely to correspond
to physiological 
sorting, but could
be related to pathological conditions where high intermolecular interaction strength induced by mutations promote the formation of
irregular, solid-like aggregates associated to degenerative deseases~\cite{BAF+18,PLJ+15}. Similar behaviors have also been observed in experiments, where overexpression of adaptor proteins responsible for mediating intermolecular interactions leads to the formation of large and irregularly shaped sorting domains~\cite{MLY+10}.  
\begin{figure*}[tb]
\begin{center}
\includegraphics[width=0.83\textwidth]{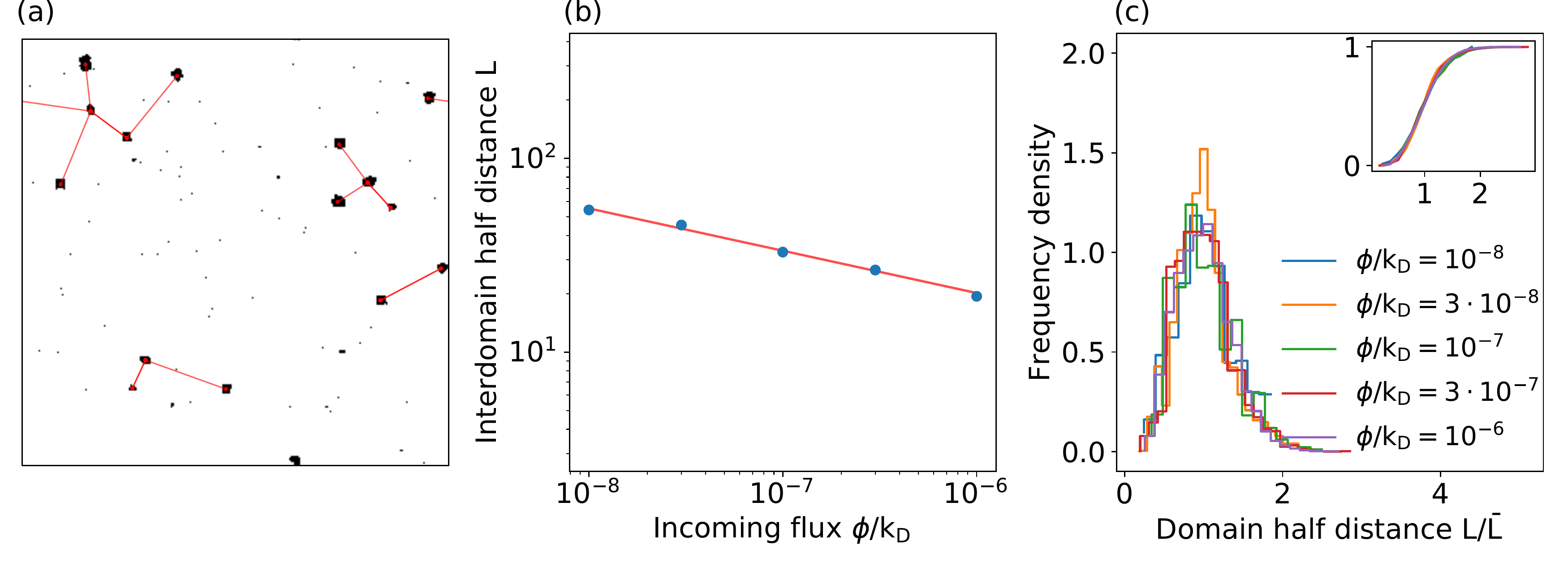}
  \end{center}
  \vspace{-0.4cm}
 \caption{(a) Nearest-neighbor distances between 
 simulated sorting
 domains are highlighted in red in a snapshot from a simulation performed with incoming flux $\phi/k_D=10^{-7}$ and interaction strength $g=10^2$. (b) Scaling of the optimal values of the average interdomain half distance.
The red line is a fit with the
power law $\phi^ {-a}$, with $a=0.23$. (c) The frequency density and cumulative frequency distribution (inset) for the rescaled half distances $L/\bar{L}$ for varying values of the incoming flux $\phi/k_D$ collapse on a single universal frequency distribution.  } 
 \label{fig:NNdist}
\end{figure*}
In summary, for varying values of the interaction strength $g$, our abstract model recapitulates two main phenomenologies. At low and intermediate values of the interaction strength $g$, the
simulated dynamics is characterized by the formation of approximately circular sorting domains via nucleation and coarsening, compatibly with the phenomenology of liquid-liquid phase separation observed in several important biological processes~\cite {BBH18,HWJ14,LPR21}. For very large $g$ instead, domain remodeling is impaired and a DLA phenomenology~\cite{BS95} is recovered, which may possibly describe the features of pathological processes. A precise characterization of the crossover between these two regimes will be the matter of future investigation.

Numerical simulations confirm the validity of
the 
scaling laws $\rho_{\rm opt}\sim \phi^a$, $\bar{n}_{\rm opt} \sim \phi^b$ and $\bar{T}_{\rm opt} \sim \phi^{-c}$, 
as the numerically obtained values $a=0.48$, $b=0.46$ and $c=0.52$ are
in 
good agreement with the theoretical predictions $a=b=c=1/2$~\cite{ZVS+21}, 
that were derived under 
simplifying assumptions.

In addition to these former results, other 
predictions
of the phenomenological theory can be 
verified numerically using the microscopic lattice-gas model. The previously exposed phenomenological theory is 
valid in the regime where
supercritical domains are well separated objects, with a well defined value of the average interdomain half distance $\bar{L}$. Since the number of supercritical domains scales
as 
$\bar{N}_d \sim \phi^{1/2}$, and 
$\pi \bar{L}^2 \bar{N}_d \approx 1$,
it is expected that $\bar{L} \sim \phi^{-1/4}$. 
This scaling law can be verified numerically 
in the following way.
First, the center of mass of each domain is computed. A critical size is determined using the operational definition given in the following Sect.~\ref{sec:criticalsize}. Domains 
with size smaller than the critical size are neglected. 
The
nearest neighbour of each domain is found 
(Fig.~\ref{fig:NNdist}(a)).
Finally,
the distances between nearest neighbors and the corresponding statistical measures
are computed. 
The numerical values of the average interdomain half distance $\bar{L}$ obtained by this method follow a scaling law $\bar{L}\sim \phi^{-d}$ with $d=0.23$, close to the theoretically predicted value $d=1/4$ (Fig.~\ref{fig:NNdist}(b)). When the mean value $\bar{L}$ is used to rescale the interdomain half distances, the corresponding frequency distributions for different values of $\phi$ collapse on a single universal distribution (Fig.~\ref{fig:NNdist}(c)).

\begin{figure}[tb]
\begin{center}
~\vskip0.16cm
\includegraphics[width=0.322\textwidth]{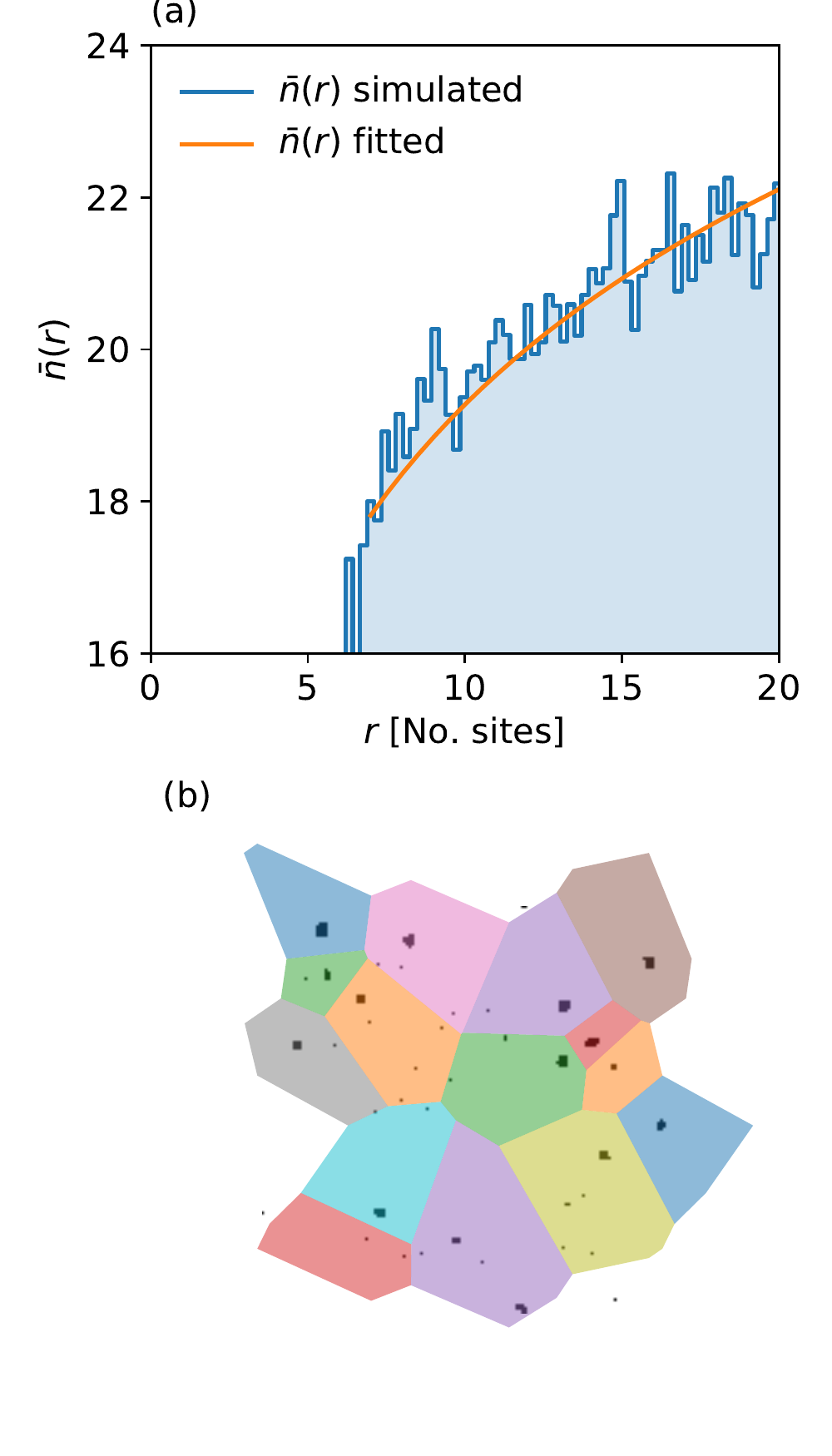}
 \end{center}
 \vspace{1.3cm}
 \caption{(a) Average density profile $\bar{n}(r)$ of the gas of free molecules at a distance $r$ from the center of supercritical domains, obtained from the simulations, and fitted with the theoretical prediction
 Eq.~\ref{eq:profile_prediction}
 ($\phi/k_D=10^{-7}$, $g=10^2$). (b) Voronoi decomposition obtained from a set of simulated supercritical sorting domains.
  (\ref{eq:profile_prediction}). }
\label{fig:densityprofile}
\end{figure}

Several results of the phenomenological theory stem from the assumption that the steady-state profile of molecule density around a sorting domain has the logarithmic form (\ref{eq:densprofile}), and from the related idea that the membrane region can be divided into  ``attraction basins'' of linear size $\sim L$ pertaining to distinct sorting domains. Given the approximate nature of these hypotheses, it is interesting to check their validity by direct numerical simulations.
A convenient 
way 
to computationally  define 
this kind of 
attraction basins
is the use of a 
Voronoi decomposition, 
which is a partition of the plane 
into
non-overlapping
regions 
according to their proximity to 
points 
of 
a given set~\cite{OBS+00}.
The two-dimensional square lattice used for the numerical simulations 
was therefore
decomposed according to the following procedure. Once all supercritical domains 
were
identified and tracked, for each time frame the center of mass of each domain was computed and
the set of these centers was
~used to partition the lattice area into Voronoi regions
(Fig.~\ref{fig:densityprofile}(b)).
Then, free
molecules belonging to each region 
were identified,
and their distance from the domain center of mass
computed.
A~direct validation of the theoretical expression
(\ref{eq:densprofile}) is computationally very demanding, as it requires building histograms of distances conditional to the radius~$R$ of a given sorting domain.  We 
studied
a slightly different quantity,
i.e. the 
average
frequency of
the 
distances of 
free
molecules from domains of 
linear
sizes 
$R$ comprised
between the critical radius $R_{\rm c}$ and the extraction radius~$R_E$: 
\begin{equation}\label{eq:nr}
 \bar{n}(r)= \int_{R_{\rm c}}^{R_E} n_R(r) N_\mathrm{st}(R) \,\mathd R
\end{equation}
for $0\leq  r \leq L$, where the theoretical model describes a density profile characterized by gas depletion in the proximity of the sorting domain.
Computing the integral in
(\ref{eq:nr}) we obtain 
\begin{equation}
    \bar{n}(r) = K_1 +K_2 \log(r),
    \label{eq:profile_prediction}
\end{equation}
where $K_1$ and $K_2$ are functions of 
the model parameters. 
If 
$p(r)\,\mathd r$ is  the
empirical
probability of finding a 
molecule
at a distance
comprised between
$r$ 
and $r+\mathd r$
from the center of mass of a domain,
then
\begin{equation}
    \bar{n}(r)=\frac{p(r)}{2\pi r}.
\end{equation}
The 
measure of $\bar{n}(r)$ obtained 
from the numerical simulations by this procedure
is in
agreement with 
a fit  of
the theoretical prediction
(Fig.~\ref{fig:densityprofile}(a))

\begin{figure}[tb]
\begin{center}
\includegraphics[width=0.38\textwidth]{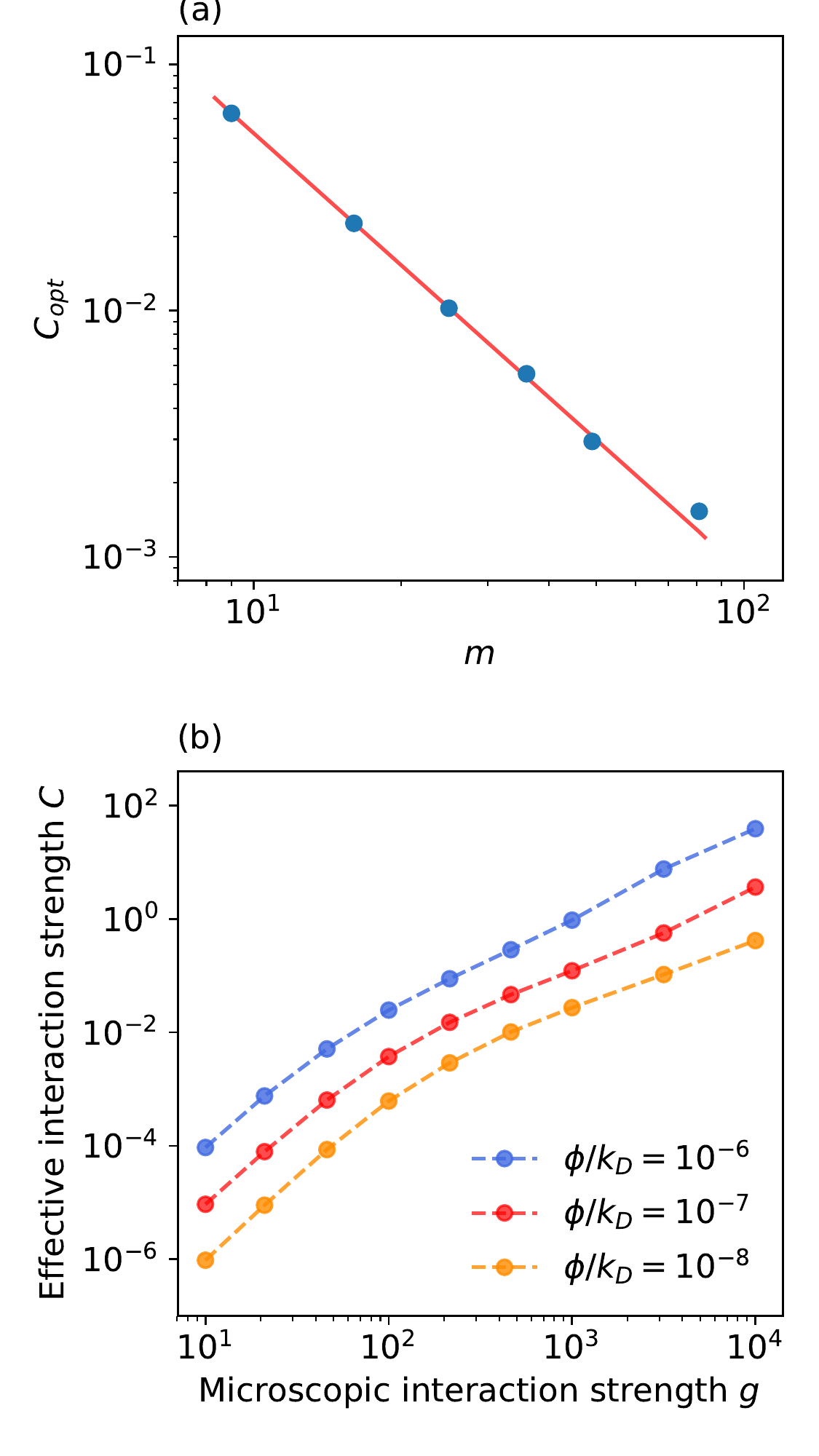}
\end{center}
\vspace{-0.5cm}
\caption{(a) Optimal effective interaction strength $C_{\rm opt}$ as a function of 
$m=A_E/A_0$, at fixed 
$\phi/k_D=10^{-6}$. The red line is a fit with the power law  
$m^{-h}$, with $h=1.8$. (b) Effective interaction strength $C$ as a function of the microscopic interaction strength~$g$, for different values of the incoming flux.}
\label{fig:Coptvsm}
\end{figure}

In the phenomenological theory, a central role is played by the dimensionless effective interaction strength $C$.
A~convenient expression for 
$C$, amenable to empirical estimation, can be obtained by inverting Eq.~\eqref{eq:cdn2} and making use of (\ref{eq:nd}) to get
\begin{equation}\label{eq:operC}
C = \frac{\phi}{m D \,\bar{n}^2},
\end{equation} 
which is a
function of directly measurable quantities, such as the incoming 
flux $\phi$ and the bulk gas density~$\bar{n}$. 
The theory predicts that the optimal value 
$C=C_\mathrm{opt}$ scales as $m^{-h}$, with $h=2$ (cf. Eq.~\ref{eq:coptimal}). 
Numerical simulations yield
the compatible value 
$h=1.8$ 
(Fig.~\ref{fig:Coptvsm}(a)).

One of the main tenets of the phenomenological theory is the existence of a well-defined critical domain size~$A_{\rm c}$, 
arising
from the balance between the mixing power of lateral diffusion and the tendency of 
sorted
molecules to aggregate. 
In the lattice-gas model, the 
tendency to aggregation is controlled by the microscopic parameter~$g$, while in the phenomenological theory, an analogous role is played by the effective interaction strength $C$.
The operational definition 
provided by
Eq.~\ref{eq:operC} 
allows to 
determine $C$ from the simulated molecule
density $ \bar{n}$ as a function 
of model parameters  
(Fig.~\ref{fig:Coptvsm}(b)).
Accordingly
with its interpretation as an effective interaction strength, $C$ is observed to be a non linear, monotonically increasing 
function of the microscopic parameter~$g$. 

The critical domain size $A_\mathrm{c}$
is a central control parameter of the molecular distillation process, but there is no simple analytical expression for it in the framework
of the phenomenological theory. 
Explicit approximate expressions for
the critical size 
can be obtained 
using
classical metastability analysis in 
quasi-equilibrium lattice-gas models (see App.~\ref{app:meta} and references therein). 
Such
an analysis 
predicts that $A_{\rm c}$ is a monotonically decreasing function of the \textit{microscopic} interaction strength between sorted molecules, 
which, however, is not 
practically measurable.  
For this reason, in the next Section 
we 
provide an operational
definition of critical size 
that
can be more
directly related to the analysis of
experimental observations.

\section{Operational definition of 
the 
critical size}\label{sec:criticalsize}
\begin{figure}[tb]
\begin{center}
\includegraphics[width=0.37\textwidth]{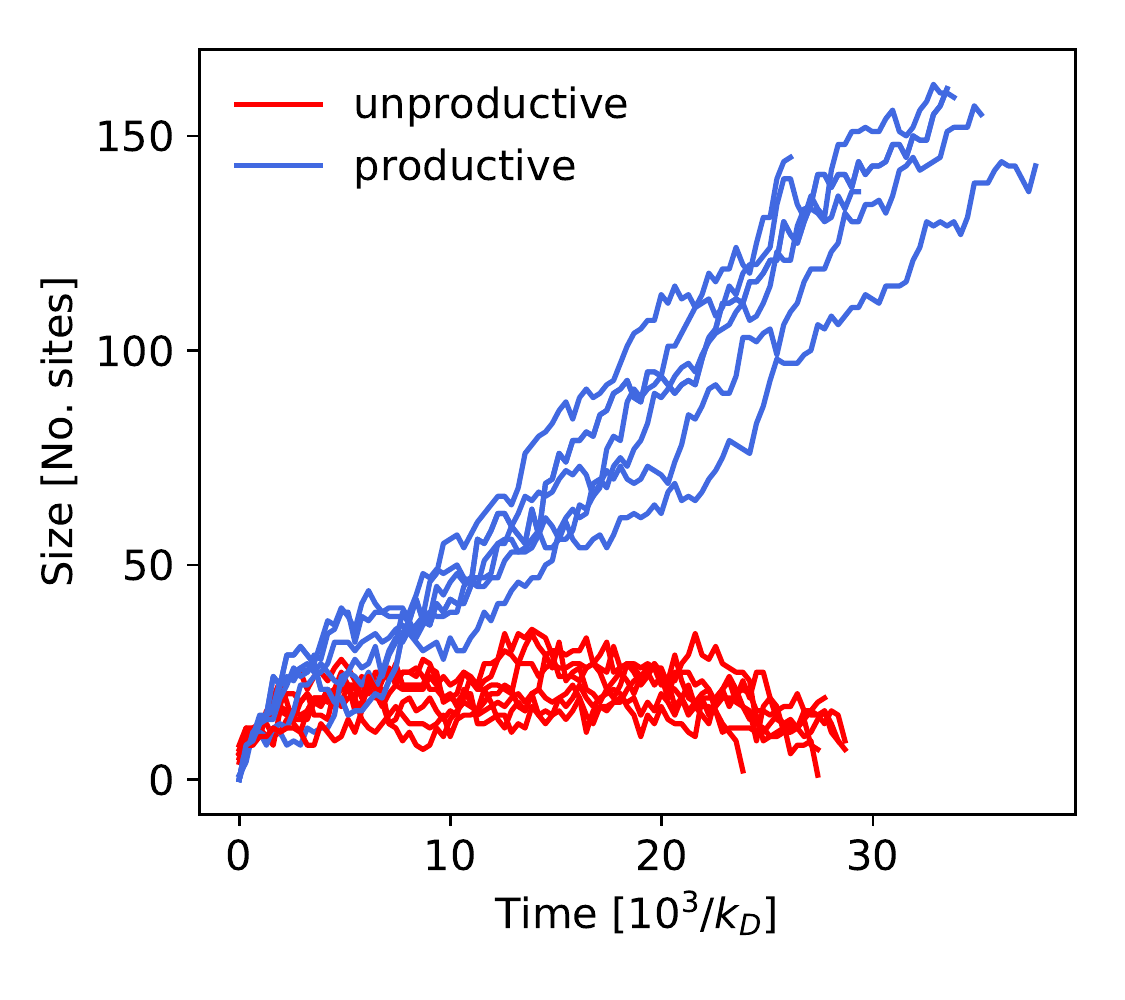}
 \end{center}
 \vspace{-0.5cm}
  \caption{Time evolution of the size of productive (blue)
  and unproductive (red) sorting domains, from numerical simulation of the lattice-gas model 
  ($\phi/k_D=10^{-6}$,  
  $g=20$). }
  \label{fig:trajectories}
\end{figure}

In experimental studies of molecular sorting, domain ``trajectories'' have been observed to fall into two classes, depending on their fate~\cite{EBV+04,AAM+13,WCM+20}: \textit{productive} trajectories, where the domain is finally extracted as a part of a lipid vesicle, and \textit{unproductive} trajectories, where the domain progressively dismantles and is ultimately dissolved.
It is worth observing here that these are properties of the domain \textit{history}, and not of its state at a given instant.
However, for simplicity, we will define 
in what follows as \textit{productive} or \textit{unproductive} domains, those that belong to productive or unproductive trajectories, respectively.
In our lattice-gas model, productive and unproductive domains can be directly distinguished by tracking their evolution in time,
and checking 
whether their trajectory ends up with an extraction event, or not~(Fig.~\ref{fig:trajectories}).
The classification into 
productive and unproductive trajectories
can be used to provide a natural, operational definition of critical size, applicable to the analysis of actual experimental data.
\begin{figure*}[tb]
\begin{center}
\includegraphics[width=1\textwidth]{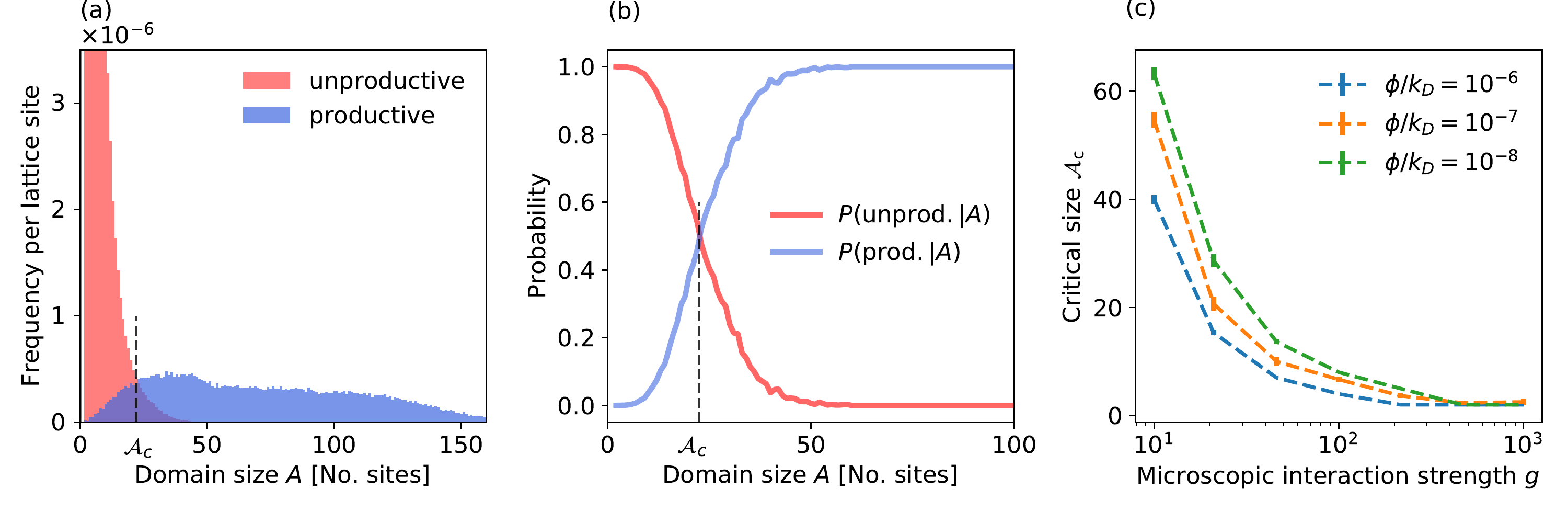}
  \end{center}
    \vskip -0.5cm
 \caption{(a) Empirical histograms 
 of
domain sizes 
 for productive (blue) and unproductive (red) domains
 obtained from 
 numerical simulations of the lattice-gas model
 ($\phi/k_D=10^{-7}$, 
 $g=20$). (b) Probability  of a domain being productive or unproductive, conditioned by its size~$A$. 
 The vertical dashed lines mark the position of the 
 critical size $\acrit$, that can be found, according to (\ref{prob3}), where the frequency of productive domains surpasses the frequency of unproductive domains (a), or~equivalently, according to (\ref{prob2}), where the conditional probability of a domain of size $A$ being productive exceeds $1/2$ (b). (c) Critical size~$\acrit$ as a function of the interaction strength~$g$ for different values of the incoming flux~$\phi/k_D$.}
\label{fig:criticalsize}
\end{figure*}
\begin{figure*}[tb]
\begin{center}
    \includegraphics[width=0.7\textwidth]{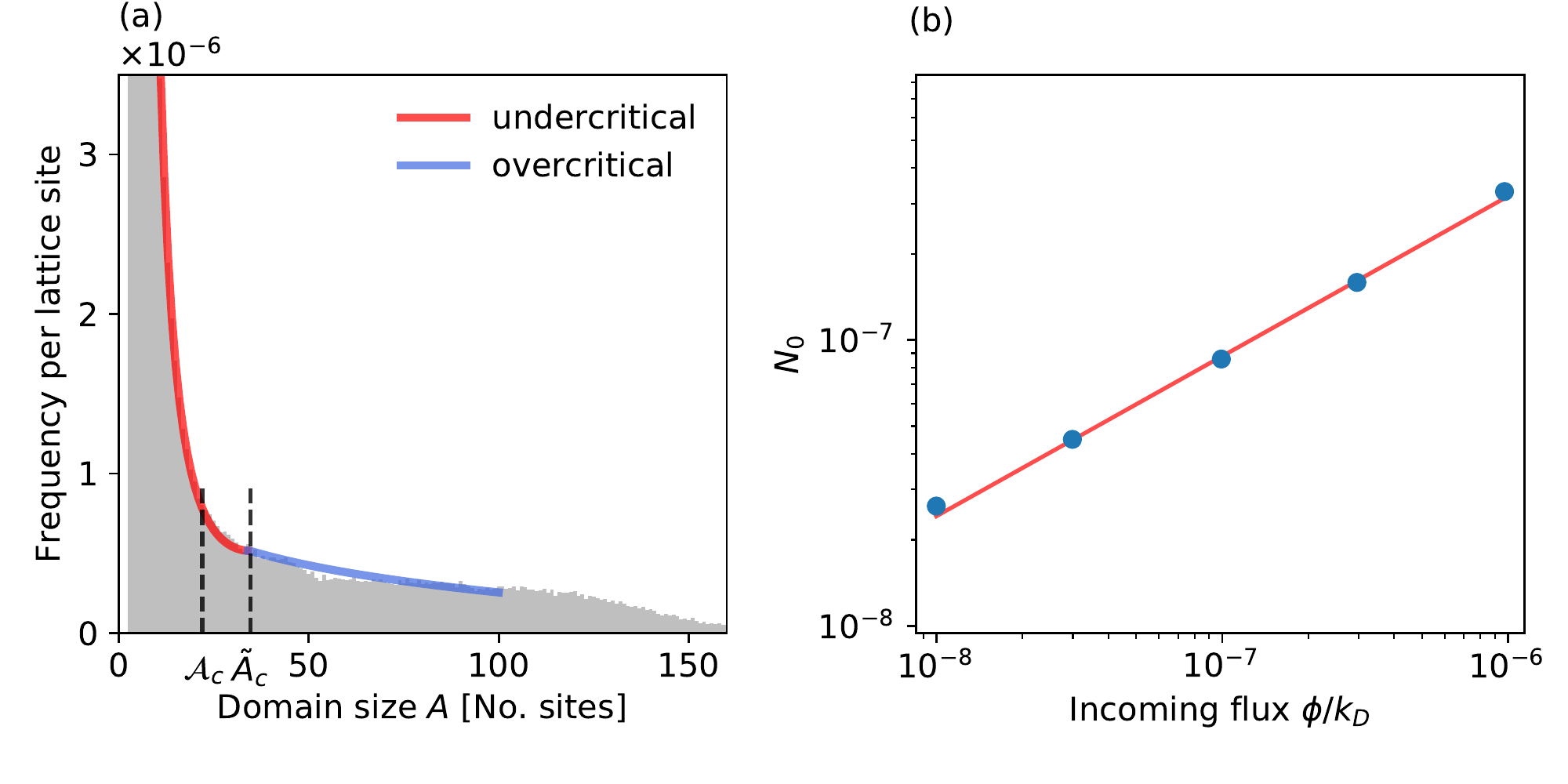}
  \end{center}
  \vskip -0.5cm
  \caption{(a) Full histogram of all domain sizes 
 ($\phi/k_D=10^{-7}$, 
 $g=20$). The lines are fits with Eq.~\ref{eq:domainsize1} (red)
  for $A<\tilde{A}_\mathrm{c}$,
 and with Eq.~\ref{eq:domainsize2} (blue) 
 for $\tilde{A}_\mathrm{c}<A<A_E$.
 The $A>A_E$ tail depends on the details of the extraction mechanism and is therefore non universal. (b) Numerical estimate of the 
 prefactor $N_0$ 
  appearing in
 Eq.~\ref{eq:domainsize2}, 
 as a function of the incoming flux $\phi/k_D$, in~the optimal region. The red line is a fit with the power-law $\phi^f$ with
 $f = 0.54$.  }
\label{fig:domainsize}
\end{figure*} Let us
define the 
`operational' critical size as 
the 
value $\acrit$
such that a domain 
of size 
$\acrit$ has 50\% probability of being productive:
\begin{equation}
P( \mathrm{prod.}|\acrit ) = \frac{1}{2},
    \label{eq:prob}
\end{equation}
(similar definitions have been adopted in previous works, see e.g. Ref.~\citenum{ryu2010numerical}).
 In terms of (joint) probability density functions (pdf's), Eq.~\ref{eq:prob} is equivalent to
 \begin{equation}
 p(\acrit,\mathrm{prod.})
 =p(\acrit,\mathrm{unprod.}),
 \label{eq:firstcrit}
 \end{equation}
 i.e., the critical size is found at
 the intersection of the joint pdf's of, respectively, productive and unproductive domain sizes.
 Under a few additional hypotheses (see App.~\ref{app:alternative_english}), Eq.~\ref{eq:prob} implies
 \begin{equation}
     P( \mathrm{prod.}|A ) 
     \geq \frac{1}{2}
     \quad\mathrm{for\ all}\quad
     A\geq\acrit
    \label{prob2}
 \end{equation}
 consistently with the phenomenological  
picture, where
smaller domains decay with high probability,
while, 
once a domain 
exceeds
the critical size, the probability that it will continue to grow up to the extraction size is larger than the probability that it will disappear.
In terms of the joint pdf's of, respectively, productive and unproductive domains, Eq.~\ref{prob2} is in its turn equivalent to the condition that
 \begin{equation}
 p(A,\mathrm{prod.})
 \geq p(A,\mathrm{unprod.})
 \quad
 \mathrm{for\ all}
 \quad
 A\geq\acrit.
     \label{prob3}
 \end{equation}
Either (\ref{prob2}) or (\ref{prob3}) can be conveniently applied to the analysis of empirical data, which are given as integer or floating-point numbers of finite precision.
The critical size $\acrit$ can thus be estimated 
either 
from 
 conditional frequencies (using  Eq.~\ref{prob2}) or from frequency histograms of domain sizes (using Eq.~\ref{prob3}), as long as productive and unproductive domains can be effectively discriminated.
As an example, in Fig.~\ref{fig:criticalsize}(a), 
$\acrit$ is found at the approximate intersection of the (joint) frequency histograms of, respectively, productive and unproductive domains.
The existence of this intersection appears to be guaranteed by the fact that $p(A,\mathrm{unprod.})$ is 
a decreasing function of $A$, while $p(A,\mathrm{prod.})$ 
is initially increasing.
Fig.~\ref{fig:criticalsize}(b) 
shows 
that the probability of a domain being productive increases with its size, while the complementary probability of being unproductive decreases.
The 
above procedure
allows to compute $\acrit$ from numerical simulations for different values of model parameters. The critical size~$\acrit$ is thus found to be 
a decreasing function 
of both
the microscopic interaction strength~$g$, and of the incoming 
molecule flux $\phi$ (Fig.~\ref{fig:criticalsize}(c)).
\begin{figure*}[tb]
\begin{center}
\includegraphics[width=1.0\textwidth]{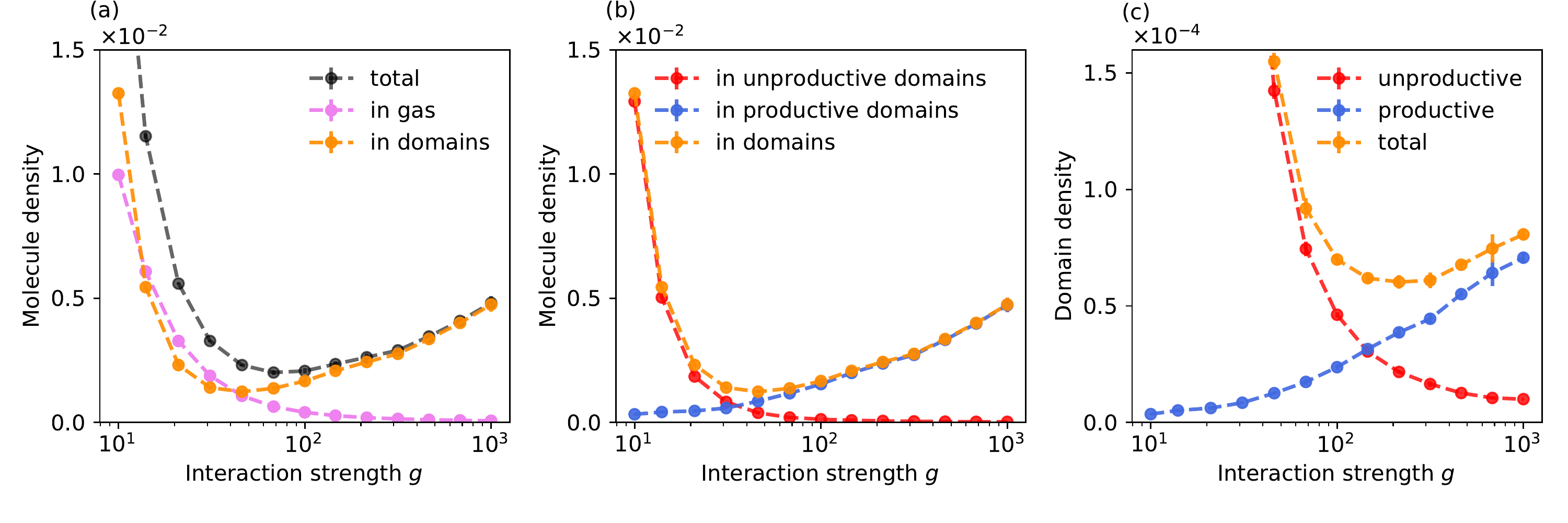}
  \end{center}
  \vskip -0.7cm
  \caption{(a) The number of free molecules per unit area decreases for increasing interaction strength~$g$ (magenta), while the  number of molecules found inside of sorting domains has an increasing trend at large~$g$ (orange). As a consequence, the total number of molecules per unit area (black) has a minimum, which marks the position of the optimal sorting regime~\cite{ZVS+21}. (b) In its turn, the number of molecules inside of sorting domains (orange) is a non-monotonic function of the interaction strength~$g$. This can be understood as follows. The number of molecules inside of unproductive domains (red) decreases with increasing interaction strength, while the number of molecules inside of productive domains (blue) increases. As a consequence, the total number of molecules found inside of sorting domains of any of the two types (orange) has a minimum close to the optimal sorting regime. (c) Similarly, the number of
  unproductive domains per unit area (red) decreases with the  interaction strength, whereas the number of productive domains (blue) increases. As a consequence, the total number of sorting domains of the two types (orange) has a minimum for intermediate interaction strength, close to the optimal sorting regime.
   Simulations performed with $\phi/k_D=10^{-8}$. The number of both productive and unproductive domains increase with increasing $\phi$ (not shown here).
}
\label{fig:numofdomains}
\end{figure*}
\begin{figure*}[tb]
\begin{center}
\includegraphics[width=1\textwidth]{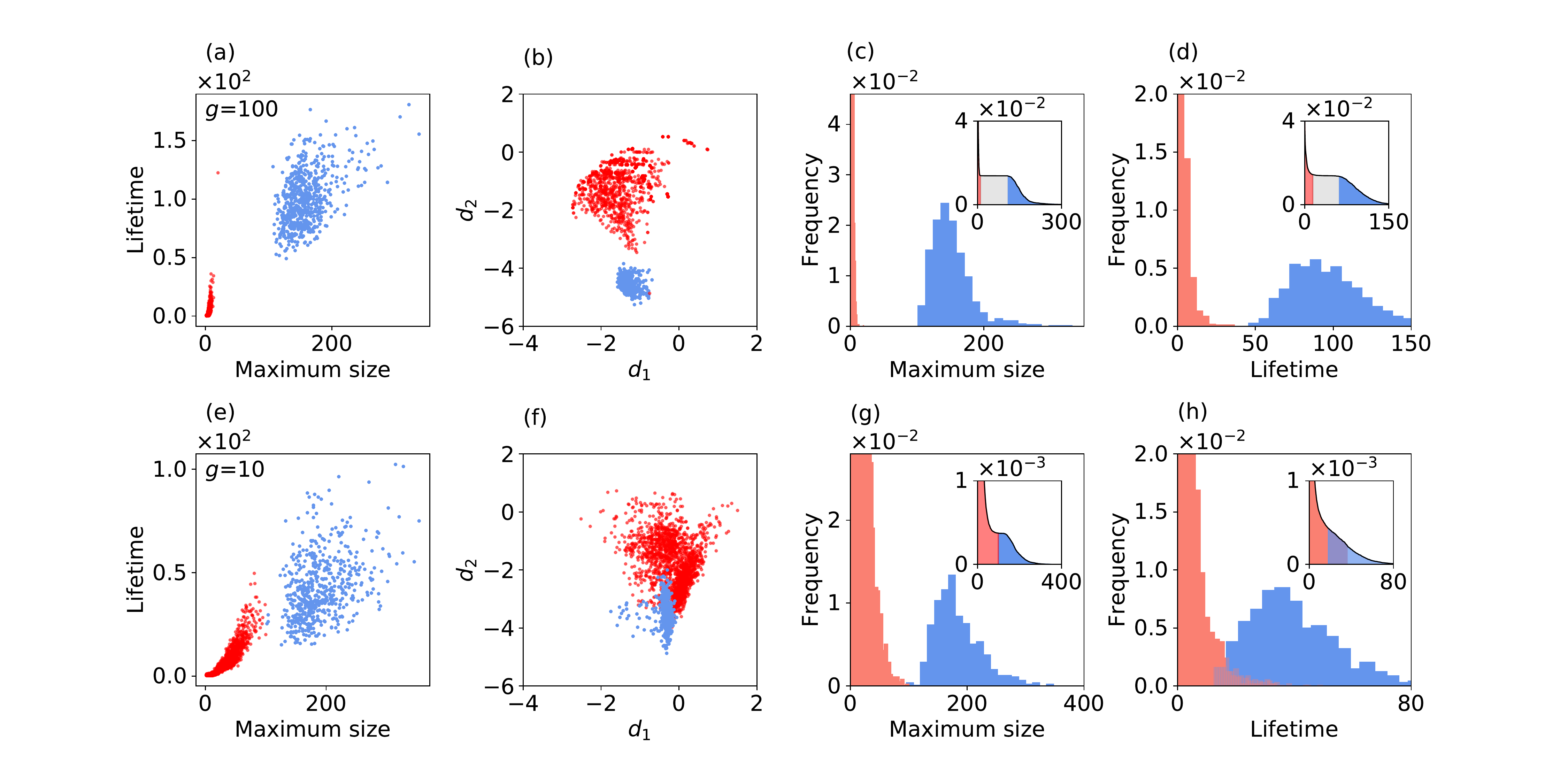}
  \end{center}
  \vskip -0.5cm
  \caption{Statistical properties of productive (blue) and unproductive (red) domains for incoming flux $\phi/k_D=10^{-6}$ and interaction strength $g=10^2$ (a-d, $5\cdot 10^4$ domain trajectories) and $g=10^1$ (e-h, $1.5\cdot 10^6$ domain trajectories),
  collected over a $3\cdot 10^6/k_D$ time interval.
  Simulated trajectories were classified into
  productive 
  and unproductive depending on whether they ended up in an extraction event, 
  or not. (a, b, e, f)
  Scatter plots of domain lifetimes vs. 
  maximum sizes (a, e) and of DASC indicators $d_1,d_2$ (b, f). 
  (c, d, g, h) frequency distributions of maximum sizes and lifetimes. Insets: complementary cumulative frequency distributions.
  Domain sizes are given as number of
  occupied lattice sites, lifetimes
  are measured in units of $10^3/k_D$.}
\label{fig:classification}
\end{figure*}

\begin{figure}[tb]
\begin{center}
\hspace{-0.5cm}\includegraphics[width=0.51
\textwidth]{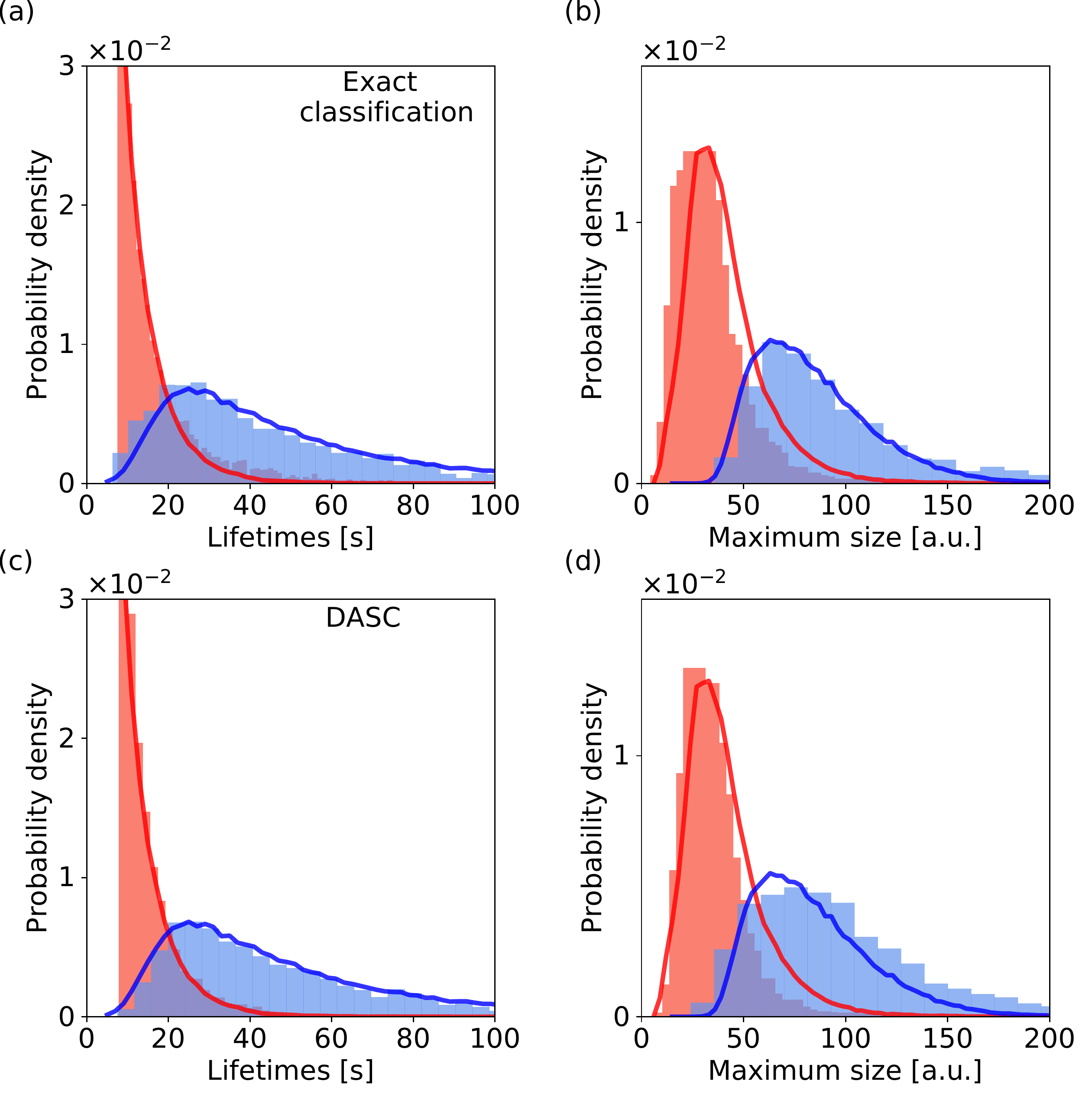}
  \end{center}
  \vskip -0.5cm
 \caption{
 Comparison 
 between the
experimental
 distributions of lifetimes (a, c) and maximum sizes (b, d) of unproductive (red lines) and productive (blue lines) domains 
 from Ref.~\citenum{WCM+20},
 Fig.~2B,C (kindly shared by Dr. Xinxin Wang),
 and 
 corresponding
 distributions obtained 
 from simulations of the lattice-gas model 
 (red and blue histograms, respectively)
 with fitted values of the model parameters
 ($g=6.5$, $\phi/k_D = 10^{-6}$) and fitted rescaling factors for lifetime and domain size units ($k_D=715\,\mathrm{s}^{-1}$, 1~lattice site\;=\;0.3~a.u.). 
 Lower cutoffs on lifetime and maximum size approximately equal to 
 the values reported in the experimental 
 data 
 were used.
 In the experiments, productive and unproductive domains were classified by DASC. In the analysis of simulated data (histograms), use was made of both the exact classification obtained directly from the simulations (a, b), and a posteriori use of DASC on the numerically generated domains (c, d), obtaining similar results.
 }
\label{fig:comparison}
\end{figure}

Having
at our disposal 
an operational definition of critical size, we are now 
in a position to check numerically the validity of theoretical predictions about
the shape of the domain size distribution.
The theory predicts functionally different forms for the number densities for the size of, respectively, subcritical and supercritical domains.
In the subcritical region, 
transient domains
continuously form and 
dissolve. This
quasi-equilibrium 
state
is approximately described by 
classical nucleation
theory~\cite{BD35,Zel43}, 
which predicts that
the 
stationary number density for domains
of size $A<A_{\rm c}$ is
\begin{equation}\label{eq:domainsize1}
N_{\rm st}^\mathrm{sub}(A) = N^\mathrm{sub}_{0}  
\mathrm{e}^{\lambda \left(A^{1/2}-A_{\rm c}^{1/2}\right)^2},
\end{equation}
where 
$\lambda$ is a constant, which is expected to be proportional to the interaction strength between sorted molecules.

For $A>A_{\rm c}$, according to 
Eq.~\ref{solutionSmolu},
the shape of the 
number density
is instead 
of the logarithmic type:
\begin{equation}\label{eq:domainsize2}
    N_{\rm st}(A)= N_{0} \log \frac{A_L}{A},   
\end{equation}
with ${N}_0 \sim \phi^{1/2}$.
By fitting the full histogram of all domain sizes with 
Eq.~\ref{eq:domainsize1} for small $A$ and with
Eq.~\ref{eq:domainsize2}
for large $A$, and by imposing the
continuity condition 
\begin{equation}
N^\mathrm{sub}_0=N_0 \log \frac{A_L}{A_{\rm c}}
\end{equation}
one obtains an estimate
$\tilde{A}_\mathrm{c}$ of the critical size $A_\mathrm{c}$
in the framework of classical nucleation theory (see Fig.~\ref{fig:domainsize}(a)).
The thus obtained value $\tilde{A}_\mathrm{c}$ is of the same order as the previously introduced  value $\acrit$,
the difference being due to the presence of a small tail of unproductive domains 
with $A>\acrit$ (see Fig.~\ref{fig:criticalsize}(a)). The definition of $\acrit$ has a clear probabilistic interpretation and is independent of phenomenological assumptions about the underlying process of domain formation. On the other hand, the estimate~$\tilde{A}_\mathrm{c}$ by the above empirical fitting procedure can be used when it is not possible to discriminate between productive and unproductive domains.

A numerical estimate of 
the prefactor $N_0$ 
for different
values of the incoming molecule flux $\phi$ gives 
$N_0\sim \phi^f$ with $f=0.54$, 
in reasonably good agreement with the theoretical value $f=1/2$ (Fig.~ \ref{fig:domainsize}(b)).

The systematic discrimination of productive and unproductive domains allows to unravel additional aspects of the phenomenology.
Optimal sorting takes place when the total number of molecules in the system 
is
minimal~\cite{ZVS+21} 
(Fig.~\ref{fig:numofdomains}(a)).
In a neighborhood of this optimal value,
one observes also a minimum 
in 
the number of molecules contained in the domains (Fig.~\ref{fig:numofdomains}(b)), and in the number of domains itself
(Fig.~\ref{fig:numofdomains}(c)).
This is a somehow
paradoxical effect, since at first sight, one would expect that a larger number of sorting domains could increase the speed of the sorting process.  Instead, sorting turns out to be most efficient precisely when the number of sorting domains is close to a minimum.  
As a matter of fact, when the interaction strength increases, the number of molecules in unproductive domains decreases, 
while the number of those in productive domains increases. As a consequence, their sum, i.e. the number of molecules in any of the two types of domains, has a minimum (Fig.~\ref{fig:numofdomains}(b)).
A similar argument applies directly to the total numbers of productive and unproductive domains: the number of unproductive domains decreases when the interaction strength increases, while the number of productive domains increases, as predicted by Eq.~\ref{eq:enned}~\footnote{Recalling also that the macroscopic interaction strength $C$ is a monotonically increasing function of the microscopic parameter~$g$ (Fig.~\ref{fig:Coptvsm}(b)).}. This leads to the appearance of an intermediate minimum in the total number of domains (Fig.~\ref{fig:numofdomains}(c)).
The 
emerging
picture is that the efficiency of the sorting process is not favored by a proliferation in the number of sorting domains: in that case, the flux of incoming molecules has to be shared among a larger number of domains, and the growth rate of individual domains is slowed down. A balance has therefore to be struck between two competing requirements: the interaction strength should be large enough to allow for easy nucleation of new sorting domains, but small enough to avoid their unnecessary proliferation. 

These theoretical predictions are compatible with former experimental work where the strength of interaction between transferrin receptors on cell plasmamembranes was experimentally controlled, and higher interaction strength was shown to induce higher rates of generation of productive sorting domains, and lower numbers of unproductive events~\cite{LAD+10}.

\section{Interpretation of experimental data}
\label{sec:classification}

The
correct classification of
productive/unproductive trajectories in 
data obtained from living cell experiments is a challenging process. 
Several approaches have been adopted. Productive trajectories can be singled out by detecting bursts in the concentration of specific molecules involved in the process of vesicle detachment, such as dynamin~\cite{FC12,GCH+14,EBV+04}. Other approaches rely on the measure of extremal properties of domain trajectories, such as the maximum size reached by domains, or their lifetime~\cite{EBV+04,LMY+09,AAM+13,HCD15,ZVS+21}, which are expected to be 
less dependent on the small-scale details of the stochastic process.
More recently, a new classification method based on a ``disassembly asymmetry score'' (DASC)~\cite{WCM+20} has been proposed. In this context,  productive and unproductive trajectories are discriminated by clustering the values of a set of statistical indicators that compare  properties of the backward and forward histories of the domains~\cite{WCM+20}.
The effectiveness of some of these approaches can be tested on numerical simulations of the lattice-gas model discussed in the previous Sections, where the
productive vs. unproductive classification can be performed exactly. 
The first two columns of Fig.~\ref{fig:classification} show scatter
plots of maximum size vs. 
lifetime (Fig.~\ref{fig:classification}(a, e)), and of the DASC indicators $d_1,d_2$~\cite{WCM+20} (Fig.~\ref{fig:classification}(b, f)),
for 
$g=10^2$ and $g=10^1$.
Different colors 
are used for 
productive (blue) and unproductive (red) trajectories.
For $g=10^2$
the two  populations 
are clearly separated,
and can be easily discriminated automatically using standard clustering methods. 
For 
$g=10^1$
 instead 
 the representative points of the two populations start to overlap, and 
 clustering methods are likely to return a certain number of erroneously classified points.
For 
$g=10^2$
the existence of two distinct populations of domain trajectories is reflected in the bimodal shape 
of the 
frequency distributions
of
maximum sizes 
and 
lifetimes 
(Fig.~\ref{fig:classification}(c, d)). This clear separation corresponds to a distinct plateau in the (complementary) cumulative frequency distribution (insets).
For $g=10^1$ 
instead (Fig.~\ref{fig:classification}(g, h)), the frequency distributions of the two populations start to overlap and the bimodal character of the two frequency distributions tends to disappear.
The loss of discriminating power takes place approximately for values of the interaction strength 
such that the critical size~$\acrit$ becomes of the order of the extraction size~$A_E$ (cf. Fig.~\ref{fig:criticalsize}(c)).

Interestingly, the model predictions for the frequency distributions of the maximum sizes and lifetimes of sorting domains are similar to those resulting from experimental observations. 
In particular, the maximum size and lifetime distributions for unproductive domains show a rapid monotonic decay, while the corresponding distributions for productive domains show a distinct maximum and a slower decaying tail (Fig.~\ref{fig:classification}(c, d, g, h)). Both of these features
have been 
observed in experiments of endocytic sorting~\cite{AAM+13,WCM+20}, where productive and unproductive domains correspond to clathrin-coated pits (CCPs) and abortive coats (ACs), respectively. (A~third population of outlier traces (OTs)~\cite{WCM+20}, characterized by short lifetimes and large sizes, likely correspond to cytoplasm-originated events~\cite{HCD15} and are not observed in the simulations.)
We looked for model parameters providing the best fit of simulated frequency distributions with data from Fig.~2B,C of Ref.~\citenum{WCM+20}, where productive and unproductive domains were classified using DASC.
By a single fit of the two parameters of the model and of two rescaling factors for the time and length scales, good agreement between simulation and experimental data was found for both the lifetime and maximum size distributions, simultaneously for both productive and unproductive domains (Fig.~\ref{fig:comparison}).
The frequency histograms obtained from the exact classification of simulated productive and unproductive domains (Fig.~\ref{fig:comparison}(a, b)) was compared with the frequency histograms obtained with the same model parameters, where however simulated domains were classified by the DASC method, yielding similar results (Fig.~\ref{fig:comparison}(c, d)).

\vskip 0.3cm
\section{Conclusions}\label{sec:conclusions}
To 
generate and
maintain 
their
internal order and guarantee 
proper physiological functioning, 
eukaryotic cells rely
on a sophisticated process 
by which 
specific biomolecules
are
sorted and concentrated on
small lipid vesicles, 
that are later delivered to appropriate membrane subregions through well-defined pathways.
A recently 
proposed
phenomenological theory of molecular sorting 
assumes that this process emerges from
the coupling of two simpler biophysical mechanisms~\cite{ZVS+21}: a) the tendency of similar molecules to phase separate into 
localized sorting
domains, and b) domain-induced membrane bending, leading
to the formation 
and ultimate detachment
of specifically 
enriched vesicles. 
A central notion of the theory of phase separation is that
only domains larger than a 
critical size 
$A_\mathrm{c}$ are able
to grow indefinitely,
while smaller domains tend to be dissolved. In combination with a contextual process of domain extraction at a larger scale $A_E > A_\mathrm{c}$, this introduces a sort of ``physical checkpoint'', such that only domains that are able to reach the ``critical mass'' $A_\mathrm{c}$ can drive extraction (distillation) events, and are thus ``productive''. 
This scenario is consistent with experimental observations where, in addition to 
``productive''
long-lived domains that grow 
into 
vesicles that are ultimately extracted from the membrane,
a large number of short-lived, small domains, which 
tend to
disassemble and ultimately disappear, is also detected. 
The existence of such a ``physical checkpoint'' is reflected in the particular shape of the size distribution for productive domains (Eq.~\ref{solutionSmolu}), which exhibits a maximum at sizes of the order of the critical size $A_\mathrm{c}$, a slowly (logarithmically) decaying intermediate region, followed by a non-universal decaying tail at scales larger than the extraction threshold $A_E$ (Fig.~\ref{fig:criticalsize}(a), blue histogram). On the other hand, the existence of a biochemical checkpoint has also been  postulated in this regard~\cite{LMY+09,AAM+13}. It~would be quite interesting to further investigate the relation between these two effects. It is worth observing here that in the actual biophysical process, a wealth of different biomolecular species takes place in the formation and stabilization of sorting domains. In the theoretical model, the complex interplay between these different species is effectively encoded into the value of the single dimensionless interaction parameter~$g$. Intriguingly, even such a highly simplified abstract model, founded on basic notions from the theory of phase separation, is able to capture relevant features of the real process. This  
yields
support to the hypothesis that 
endocytic
sorting 
is driven by an
underlying phase separation process.

We have here considered a spatially homogeneous probability of nucleation of sorting domains. It has been observed however that nucleation events may cluster in ``hotspots'' or ``nucleation organizers''~\cite{NAM+11}. The origin of such hotsposts is an interesting open question, that deserves to be investigated in the framework of phase separation theory.

~

\acknowledgments
We gratefully acknowledge useful discussions with Igor Kolokolov, Vladimir Lebedev, 
Guido Serini, Carlo Campa and Roland Wedlich-S\"{o}ldner. 
We thank Xinxin Wang, Sandra Schmid and Gaudenz Danuser for kindly sharing their data  
and for their insightful observations.  Numerical calculations
were
made possible by a CINECA-INFN agreement providing access to CINECA high-performing computing resources.
\vskip -0.3cm
\appendix
\section{Rate of supercritical domain production}\label{app:islandmodel}
The rate at which supercritical domains are generated in the non-equilibrium driven stationary state of the lattice model of molecular sorting is assumed, in Eq.\eqref{eq:cdn2}, to be proportional to the square of the free molecule density $\bar{n}$. Here we provide a justification for this assumption, based on a simplified mean-field model of monomer aggregation. 

Consider a model of domain formation by means of monomer attachment and detachment, and suppose 
that there 
exists a threshold area value $A_{\rm c}$ above which monomer detachment from domains is not possible, and clusters grow irreversibly. This way, the existence of a critical size in the system is artificially reproduced. Let~us call $n_A$ the number density per unit surface of domains of area $A$, and $N_{+}$ the number density per unit surface of domains with $A> A_{\rm c}$. The incoming flux of monomers is $\phi$. The set of mean-field Smoluchowski equations for this model is:
\begin{widetext}
\begin{subequations}
\begin{align}
\frac{\mathd n_{1}}{\mathd t}	 &=	- 2 c_1 n_{1}^{2}+    b_2 n_{2}- n_{1}\sum_{A=2}^{A_{\rm c}} c_A n_{A}+ \sum_{A=2}^{A_{\rm c}}  b_A n_{A}- c_+ n_{1}N_{+} +\phi \\
\frac{\mathd n_{A}}{\mathd t}	 & =	c_{A-1} n_{1}n_{A-1}-  b_A n_{A}-  c_A n_{1} n_{A} + b_{A+1} n_{A+1},\qquad2\leq A\leq A_{\rm c}\\
\frac{\mathd N_{+}}{\mathd t} &	= c_{A_{\rm c}} n_{1}n_{A_{\rm c}}
\end{align}
\end{subequations}
\end{widetext}
where  $c_A$ (and,  respectively, $b_A$) are dimensional coefficients representing the attachment 
(detachment) rates of monomers on (from) domains of area $A$, and $b_A=0$ for $A>A_\mathrm{c}$. According to reaction rate theory \cite{krapivsky2010kinetic}, in two dimensions, the effective reaction rate of two domains is proportional to the sum of their diffusion constants. In the approximation where only monomers can move (extended domains  
being
much slower, as 
their diffusivity decreases with 
size 
as $A^{-3/2}$), the effective aggregation rates $c_A$ become 
independent of~$A$ and 
proportional to the diffusivity $D$ of a monomer. It is also important to notice that dimers can split with a rate proportional to $D/g$ per molecule, i.e. $b_2 = b/g$. Summing over the areas $2\leq A\leq A_{\rm c}$ to obtain an equation for $N_{-}=\sum_{A=2}^{A_{\rm c}}n_{A}$, we find
\begin{subequations}
\begin{align}
\frac{\mathd N_{-}}{\mathd t}	 & =	c\, n_{1}^{2} -  c\, n_{1}n_{A_{\rm c}}  - g^{-1} b\, n_2 \\
\frac{\mathd N_{+}}{\mathd t} & =	 c\, n_{1}n_{A_{\rm c}}.
\end{align}
\end{subequations}
The stationary condition $\mathd N_{-}/\mathd t=0$ for the subcritical domains implies 
$\mathd N_{+}/\mathd t = c\, n_{1}^{2} - g^{-1} b\, n_2$. In order for $N_{-}$ to be approximately constant with a non-zero production of supercritical domains, 
the second term 
must be
subdominant already at moderately large values of~$g$. One can then conclude  that the net production of supercritical domains $N_{+}$ is well approximated by the equation $\mathd N_{+}/\mathd t \approx c\, n_{1}^{2}$. The quantity $N_+$ 
corresponds
to the number density $N_d$ of  supercritical domains used in the main text, 
thus qualitatively justifying Eq.~\ref{eq:cdn2}.

\section{Metastability in lattice-gas models}\label{app:meta}
Assuming 
that 
 in a quasi-equilibrium condition the molecule gas density outside of a growing domain in a lattice-gas model follows the Gibbs-Thomson relation $n_0 =n(R) = n_\infty(1+\sigma/R)$ \cite{ryu2010numerical}, where $\sigma$ is the line tension of the domain, using Eq.~\ref{eq:dAdt} we get
\begin{equation}
\dot{R}
=\frac{A_{0}D}{\log\left(L/R\right)}\left(R-\frac{\sigma n_{\infty}}{\bar{n}-n_{\infty}}\right)\frac{\bar{n}-n_{\infty}}{R^{2}}. 
\end{equation}
  Therefore the critical value of the domain radius is
\begin{equation}\label{rc_noneq}
R_{\rm c} = \frac{\sigma n_\infty}{\bar{n}-n_\infty}.
\end{equation}
This is a non-equilibrium result, in which $\bar{n}$ and $n_\infty$ represent respectively the bulk average density of the molecule gas and the equilibrium density of the gas at the interface with a large flat domain.

These two quantities, together with $\sigma$, can be easily estimated at equilibrium in a lattice-gas model. Consider an equilibrium  lattice-gas model with a chemical potential $\mu$, and let $\epsilon>0$ be the energy gain due to the attractive interaction between two molecules occupying nearest-neighboring sites of the lattice.
The energy function of the equilibrium lattice-gas system takes the form  
\begin{equation}
E(\eta) = \mu \sum_i \eta_i -\epsilon \sum_{\langle i,j\rangle}\eta_i \eta_j,
\end{equation} 
 where $\eta=\{\eta_i\}$ with $\eta_i\in\{0,1\}$ for $i=1,\dots, \mathcal{N}$ is a binary  configuration representing the presence or absence of molecules on lattice sites. 
According to the dynamic viewpoint of Ref.~\citenum{HOS00}, the expression 
\begin{equation}
R_{\rm c} \approx  \frac{ \epsilon}{z\,\epsilon - 2\mu},
\end{equation} 
with $z$ the number of nearest neighbors of a given site, is obtained imposing a local equilibrium condition between the probability of growing and that of shrinking. In a mean-field equilibrium picture, the chemical potential is related to the average total density $\bar{n}_{\rm eq}= {\rm e}^{-\beta\mu}$ of free molecules in a supersaturated system. At the condensation point $\mu=z\epsilon/2$, the average total density is equal to the saturation density ${n}_\infty \approx {\rm e}^{-\beta z\epsilon/2}$, which is the molecule density of a gas phase in equilibrium with a liquid phase (with flat interface). 
In terms of these quantities, the critical domain radius becomes
\begin{equation}
R_{c}\approx \frac{\beta\epsilon}{2\,\log(\bar{n}_{\rm eq}/{n}_{\infty})} \approx	\frac{\beta\epsilon {n}_\infty}{2 \,(\bar{n}_{\rm eq} - n_\infty)}\label{eq:rc_eq}
\end{equation}
close to the condensation point. The expression is formally equivalent to Eq.~\ref{rc_noneq} if we identify $\sigma = \beta \epsilon/2$. Given two configurations $\eta, \eta'$ of the dynamic lattice-gas model, the detailed balance condition implies  
\begin{equation}\label{eq:dbLG}
\frac{W\left(\eta \to \eta' \right)}{W\left(\eta' \to \eta \right)} = \frac{P_{\rm eq}(\eta' )}{P_{\rm eq}(\eta )}.
\end{equation} 
Focusing on the transition $\eta \to \eta'$, in which a dimer fragments into two monomers as a consequence of one of them hopping away,
and since $z=4$ for a square lattice,
the previous relation implies 
$\beta \,\epsilon = \log{g}$, and consequently $n_{\infty} \approx e^{-2\beta\epsilon} = g^{-2}$.  Moreover, in a lattice gas at equilibrium, the average density $\bar{n}_{\rm eq}$ of supersaturated gas is fixed by the chemical potential and is independent of the microscopic interaction strength. Therefore, from \eqref{eq:rc_eq} the critical radius 
is seen to be 
a monotonically decreasing function of~$g$. 

In the non-equilibrium stationary state relevant to the description of molecular sorting, $\bar{n}$ is numerically observed to be a decreasing function of $g$, however its decrease is slower 
than  
the decrease of
$n_\infty$, since $\bar{n}$ 
is sustained by
the constant molecular influx~$\phi$. This way, from Eq.~\ref{rc_noneq} the critical radius $R_{\rm c}$ is seen to be 
a monotonically decreasing function of $g$ also in the 
non-equilibrium case of interest.
\section{Alternative definitions of critical size}\label{app:alternative_english}
Maintaining the notations used in the main text, let
us define the empirical critical size as the value $\acrit$ such that 
\begin{equation}
P( \mathrm{prod.}|\acrit ) = \frac{1}{2}.
    \label{prob}
\end{equation}
This
value 
is well defined
if 
$P( \mathrm{prod.}|A)$
is a continuous
function which tends to 0 for $A\rightarrow0$ and to 1 for $A\rightarrow \infty$. 
Equivalently, 
$\acrit$ can be defined as the
solution of
\begin{eqnarray}
  p (A, \mathrm{prod} .) & = & p
  (A, \tmop{unprod} .)  \label{eq:critsize2}
\end{eqnarray}
since (\ref{eq:critsize2}) can be rewritten as
\begin{eqnarray*}
p(A,\mathrm{prod.})&=& p(A)-p(A,\mathrm{prod.})
\end{eqnarray*}
yielding
\begin{eqnarray*}
 P (\mathrm{prod.} |A)  =  \frac{p (A, \mathrm{prod.}) }{p (A) } & = & \frac{1}{2}.
 \end{eqnarray*}
If $p(A, \tmop{unprod.})$ is a decreasing function of $A$,  and $p (A, \tmop{prod.})$ is an increasing function of $A$ in a right neighborhood of 0 (as the simulations suggest, see e.g. Fig.~\ref{fig:criticalsize}), one can easily show that $P(\mathrm{prod.}|A_1\geqslant A \geqslant \acrit)$ is a non decreasing function of $A_1$ by directly computing its derivative with respect to that variable. Then, for all $A_1\geqslant \acrit$ one has:
\begin{eqnarray*}
  P (\tmop{prod.} |A \geqslant \acrit) 
  & \geqslant & P (\tmop{prod.} |A_1 \geqslant A \geqslant \acrit)\\
  & \geqslant &  P (\tmop{prod.} |A = \acrit)  =  \frac{1}{2}\,.
\end{eqnarray*}

\end{document}